\documentclass[12pt]{article}
\usepackage[margin=1in]{geometry}
\usepackage{amsmath}
\usepackage[dvips]{graphicx}
\usepackage{epsf}
\usepackage{amssymb}
\usepackage{color}

\newcommand{\be}{\begin{equation}}
\newcommand{\ee}{\end{equation}}
\newcommand{\bea}{\begin{eqnarray}}
\newcommand{\eea}{\end{eqnarray}}
\newcommand{\barr}{\begin{array}}
\newcommand{\earr}{\end{array}}

\def\bk{{\bf k}}

\def\CT{{\cal T}}

\def\half{\frac{1}{2}}

\begin{document}

\newcommand{\z}[1]{\zeta({\bf k}_{#1})}
\newcommand{\zg}[1]{\zeta_1({\bf k}_{#1})}

\thispagestyle{empty}

\vspace*{0.3in}

\begin{center}
{\huge \bf Large Nonlocal Non-Gaussianity from a Curvaton Brane}

\vspace*{0.5in} {Yi-Fu Cai\footnote{caiyf@mail.ihep.ac.cn}}
\\[.3in]
{\em
 Department of Physics, Arizona State University,\\ Tempe, AZ, 85287-1504, USA
\vspace{0.3in} }

{\em
 Institute of High Energy Physics, Chinese Academy of Sciences,\\
 P.O. Box 918-4, Beijing 100049, China
\vspace{0.3in} }

\vspace*{0.5in} {Yi Wang\footnote{wangyi@hep.physics.mcgill.ca}}
\\[.3in]
{\em
 Physics Department, McGill University,\\ Montreal, H3A2T8, Canada
\vspace{0.3in} }
\end{center}

\begin{center}
{\bf Abstract}
\end{center}
\noindent

We use a generalized $\delta N$ formalism to study the generation of the primordial
curvature perturbation in the curvaton brane scenario inspired by
stringy compactifications. We note that the non-Gaussian features,
especially the trispectra, crucially depend on the decay mechanism
in a general curvaton scenario. Specifically, we study the
bispectra and trispectra of the curvaton brane model in detail to
illustrate the importance of curvaton decay in generating
nonlinear fluctuations. When the curvaton brane moves
non-relativistically during inflation, the shape of
non-Gaussianity is local, but the corresponding size is different
from that in the standard curvaton scenario. When the curvaton
brane moves relativistically in inflationary stage, the shape of
non-Gaussianity is of equilateral type.

\vfill
\newpage
\setcounter{page}{1}

\section{Introduction}

As an alternative to inflation, the curvaton model
\cite{Lyth:2001nq} is a well-motivated proposal for explaining the
observed scale-invariant primordial density perturbation in the
framework of inflation (and some pioneering works on this
mechanism have been considered in Refs. \cite{Mollerach:1989hu,
Linde:1996gt, Enqvist:2001zp}). This model is based on the
inflationary scenario of multiple fields, in which two fields are
required at least. In this model the universe is composed of
radiation decayed from the inflaton and the curvaton field. During
inflation, the curvaton is subdominant and provides entropy
(isocurvature) perturbation, and afterwards, the entropy
perturbation can be converted to curvature perturbation as long as
the curvaton decays into radiation before primordial
nucleosynthesis. After the curvaton decays, the universe enters
the standard thermal history, and then the primordial curvature
perturbation leads to the formation of the large scale structure
of our universe\cite{Lyth:2002my, Lyth:2003ip}.

Recently, a curvaton scenario realized in the frame of stringy
inflationary models was presented in Ref. \cite{Li:2008fma}. The
idea of brane inflation proposed in \cite{Dvali:1998pa}, is
recently successfully realized by virtue of warped
compactifications\cite{Kachru:2003si}. In the stringy landscape a
number of probe branes are allowed to move in warped throats, and
under the description of supergravity (SUGRA) their dynamics is
described by a Dirac-Born-Infeld (DBI) action in the relativistic
limit. When applied to cosmology, it was found that a
relativistically moving D3-brane in a deformed AdS throat is able
to drive inflation without slow-roll at early
universe\cite{Silverstein:2003hf, Chen:2004gc}. This idea was
later extended into the scenario of multiple brane inflation
appeared in Refs. \cite{Cai:2008if, Cai:2009hw} (see
\cite{Cline:2005ty, Piao:2002vf} for earlier studies under
slow-roll approximation). A distinguish feature of the model in
this type is that a large positive non-local non-Gaussianity can
be obtained due to an enhancement of a small sound speed for the
perturbation\cite{Alishahiha:2004eh, Chen:2006nt}, which is in
contrast to the prediction of a canonical single-field inflation
model\cite{Maldacena:2002vr}. Established on the scenario of
multiple brane inflation\cite{Cai:2008if, Cai:2009hw} and
motivated by the model of {\it spinflation} \cite{Easson:2007dh},
an explicit construction of the curvaton brane model was recently
realized in Ref. \cite{Zhang:2009gw} in which angular degrees of
freedom had been introduced to perform a relativistic rotation in
a warped throat.


The investigation of non-Gaussianities is crucial in
distinguishing the curvaton scenario from usual inflation models.
A key prediction of a general curvaton model is that a sizable
local non-Gaussianity can be obtained of which the value is mainly
decided by the occupation of curvaton energy $\Omega_{\chi}$ after
inflation characterized by a transfer efficiency parameter
$$r=\frac{3\Omega_{\chi}}{4-\Omega_{\chi}}$$
which is usually calculated on the hypersurface of the curvaton
decay. As a consequence, the decay mechanism of a curvaton field
plays an important role in determining the detailed information of
primordial perturbation seeded by the curvaton field
\cite{Sasaki:2006kq}. It was pointed out in Ref.
\cite{Sasaki:2006kq} that at third order, the primordial curvature
perturbation under the assumption of sudden curvaton decay is
nicely consistent with that obtained in the case of a
non-instantaneous decay in the limits of small $r$ and
$r\rightarrow1$; however, their bispectra and trispectra deviate
from the others in the middle band of the allowed value regime of
$r$. The implications of the non-Gaussian features of curvature
perturbation on the curvaton scenario was studied in
\cite{Huang:2008ze, Huang:2008bg, Gong:2009dh}.

In the present paper we extend the investigation of Ref.
\cite{Sasaki:2006kq} relating non-Gaussian features of primordial
curvature perturbation to the curvaton decay mechanism into a more
general curvaton scenario. Specifically, we consider a model of
curvaton brane. We study the generation of curvature perturbation
and non-Gaussianity originating from the vacuum fluctuations on
the curvaton brane in cases of a sudden decay on the slice of
uniform curvaton density, and then compare their observational
signatures with the standard curvaton model to illustrate the
significance of the decay process of the curvaton field. Moreover,
since the model of curvaton brane involves a sound speed parameter
which describes the propagation of the curvaton fluctuations
during inflation, it has interesting implications to cosmological
observations as well. Note that in our scenario, the inflaton
field can be realized by any string realization of inflation which
is independent of the curvaton field. Thanks to the virtue of
curvaton scenario, the detailed realization of inflation does not
affect the calculation for curvaton perturbation, as long as the
inflaton perturbation itself is negligible.

Our paper is organized as follows. In Section 2, we point out that
the non-Gaussian features in a curvaton mechanism crucially depend
on the curvaton decay mechanism. We suggest the curvaton could
decay on a slice of uniform curvaton density, and non-Gaussian
fluctuation evaluated on this slice is mainly determined by a
combination of the equation-of-state (EoS) of the curvaton and a
generalized transfer efficiency parameter which will be shown in
the main text. In Section 3, we study the model of curvaton brane
as a specific example. We firstly describe the background
evolution of a curvaton brane model in relativistic limit
throughout the inflationary period and the process of curvaton
decay. Afterwards, we develop an analytic investigation of the
curvature perturbation and the non-Gaussianity generated in this
model by solving the perturbation equation of the curvaton brane
order by order. In Section 4 we consider a class of plausible
curvaton brane dominant eras and derive the nonlinearity
parameters characterizing non-Gaussian distribution of the
primordial curvature perturbation in these cases respectively.
Finally, we present concluding remarks in Section 5.

\section{Curvaton decay and Non-Gaussianities}

In this section, we first briefly review the standard calculation
of bispectrum and trispectrum for curvaton mechanism, with the
assumption that the decay hypersurface of the curvaton field is a
uniform total energy density slice. Afterwards, we show that these
non-Gaussian features behave different under different curvaton
decay mechanisms. Especially, when the curvaton decay hypersurface
is a uniform curvaton energy density surface, the trispectrum is
significantly modified. We then apply our result to a curvaton
model realized by a non-relativistic moving brane in a warped
throat and study its cosmological implications.

\subsection{Curvaton decay at uniform total energy density}

The non-Gaussian features for curvaton decay at uniform total
energy density slice is well studied in the literature. Here we
briefly review the approach in \cite{Sasaki:2006kq}.

In the local ansatz of non-Gaussianity, the curvature perturbation
can be expanded order by order as follows,
\begin{align}\label{zetalocal}
  \zeta(x)=\zeta_1(x)+\frac{3}{5}f_{NL}\zeta_1^2(x)+\frac{9}{25}g_{NL}\zeta_1^3(x)
  +{\cal O}(\zeta_1^4)
  =\sum_{n=1}^{\infty}\frac{\zeta_n(x)}{n!}
  ~,
\end{align}
where $\zeta_1$ is the Gaussian fluctuation, and $\zeta_n$ are the
non-Gaussian components of order $\zeta_1^n$. The relation between
$\zeta_n$ and the non-Gaussian parameters yields the following
non-Gaussian estimators in a general curvaton model,
\begin{gather}\label{estimator}
  f_{NL}=\frac{5}{6}\frac{\zeta_2}{\zeta_1^2}~,\qquad g_{NL}=\frac{25}{54}\frac{\zeta_3}{\zeta_1^3}~.
\end{gather}
The correlation functions are defined as
\begin{gather}
  \langle \z{1}\z{2} \rangle = (2\pi)^3 P(k_1)\delta^3(\sum_{n=1}^2 \bk_n)~, \nonumber\\
  \langle \z{1}\z{2}\z{3} \rangle = (2\pi)^3 B(\bk_1,\bk_2,\bk_3)\delta^3(\sum_{n=1}^3 \bk_n)~,\nonumber\\
  \langle \z{1}\z{2}\z{3}\z{4} \rangle = (2\pi)^3 T(\bk_1,\bk_2,\bk_3,\bk_4)\delta^3(\sum_{n=1}^4 \bk_n)~,\label{def234}
\end{gather}
where $P(k_1)$ is related to the dimensionless power spectrum in
form of
\begin{equation}
  {\cal P}_\zeta(k_1)\equiv \frac{k^3}{2\pi^2}P(k_1)~.
\end{equation}
To insert the ansatz \eqref{zetalocal} into Eq. \eqref{def234},
one can relate the bispectrum $B$ and the trispectrum $T$ with $P$
as follows,
\begin{gather}\label{shapeB}
  B(\bk_1,\bk_2,\bk_3)=\frac{6}{5}f_{NL}
  \left[P(k_1)P(k_2)+2{~\rm perm.}\right]~.
  \\
  T(\bk_1,\bk_2,\bk_3,\bk_4)=
  \frac{54}{25}g_{NL}\left[ P(k_1)P(k_2)P(k_3) +3 {~\rm perm.} \right] \nonumber\\
  +\tau_{NL} \left[ P(k_1)P(k_2)P(|\bk_1+\bk_3|)+11{~\rm perm.}\right]~. \label{shapeT}
\end{gather}

On super-Hubble scales, the curvature perturbation on uniform
density slice can be written as
\begin{equation}\label{deltaN}
  \zeta_i(x)=\delta N(x)+\frac{1}{3}\int_{\bar\rho_i(t)}^{\rho_i(x)}\frac{d\tilde\rho_i}{\tilde\rho_i+P_i(\tilde\rho_i)}~,
\end{equation}
where the subscript $i$ denotes either inflaton, curvaton or total
energy density. As there is no interaction between inflaton and
curvaton, the inflaton perturbation, which will be transferred
into radiation after inflaton decay, $\zeta_r$, and the curvaton
perturbation $\zeta_\chi$ are both conserved on super-Hubble
scales.

As a first step, one need to write down the relation between the
curvaton fluctuation $\delta\chi$ and its curvature perturbation.
Choosing the spatial flat slice, Eq. \eqref{deltaN} for curvaton
becomes
\begin{equation}\label{deltaNchi}
  \rho_\chi=\bar\rho_\chi e^{3\zeta_\chi}~,
\end{equation}
in the neighborhood of the curvaton decay hypersurface.

Consider the curvaton perturbation that is initially originated
from vacuum fluctuations inside the Hubble horizon. These
perturbation modes could satisfy a Gaussian distribution at the
Hubble exit. For a curvaton brane, this assumption applies when
the brane is moving non-relativistically. In this case, we have
\begin{equation}
  \chi_*=\bar\chi_*+\delta_1\chi_*~.
\end{equation}
This Hubble-crossing value can be related to the initial amplitude
of curvaton oscillation, which takes the form
\begin{equation}
  g(\chi_*)=g(\bar\chi_*+\delta_1\chi_*)=\bar{g} +\sum_{n=1}^{\infty}\frac{g^{(n)}}{n!}\left(\frac{\delta_1\chi}{g'}\right)^n ~.
\end{equation}
The detailed form of $g(\chi_*)$ is a model-dependent function
which is determined by the explicit potential of the curvaton
field in the non-relativistic limit. For example, if the curvaton
potential is quadratic all the way until curvaton decay, then
$g(\chi_*)\propto \chi_*$. In this case, the energy density can be
written as
\begin{equation}\label{curvatondensity}
  \rho_\chi=\frac{1}{2}m^2 g^2 = \frac{1}{2}m^2 \left[ \bar{g} +\sum_{n=1}^{\infty}\frac{g^{(n)}}{n!}\left(\frac{\delta_1\chi}{g'}\right)^n
  \right]^2~.
\end{equation}
Therefore, the combination of Eqs. \eqref{deltaNchi} and
\eqref{curvatondensity} leads to
\begin{gather}
  \zeta_{\chi1}=\frac{2}{3}\frac{\delta_1\chi}{\bar\chi}~, \qquad
  \zeta_{\chi2}=\frac{3}{2}\left(-1+\frac{gg''}{g'^{2}}\right)\zeta_{\chi1}^2~, \qquad
  \zeta_{\chi3}=\frac{9}{4}\left(2-3\frac{gg''}{g'^{2}}+\frac{g^2g'''}{g'^{3}}\right)\zeta_{\chi1}^3~.
\end{gather}

A second step of the calculation is to relate $\zeta_\chi$ to
$\zeta$. In the sudden decay approximation, the relation is quite
simple and can be computed analytically. However, we also note
that this relation crucially depends on the curvaton decay
mechanism, which will be discussed in the next subsection. In this
section, we assume the curvaton decays on a uniform total density
hypersurface $H=\Gamma$, where $\Gamma$ is the decay rate of the
curvaton. This can be realized in the framework of particle
physics, namely the curvaton has a ``life time'' and then decays
to particles through their coupling terms. Then on the curvaton
decay hypersurface we have
\begin{equation}
  \rho_r+\rho_\chi=\bar\rho~.
\end{equation}
In a usual case the curvaton is oscillating around its vacuum
before its decay. Thus one obtains a pressureless field fluid on
average with $P_\chi=0$ for curvaton, and $P_r=\rho_r/3$ for the
radiation from inflaton. Making use of Eq. \eqref{deltaN}, one can
get
\begin{equation}
  \rho_r=\bar\rho_r e^{4(\zeta_r-\zeta)}~,\qquad \rho_\chi=\bar\rho_\chi e^{3(\zeta_\chi-\zeta)}~.
\end{equation}
As a consequence, $\zeta$ and $\zeta_\chi$ are related on the
decay hypersurface as follows
\begin{equation}\label{totalenergymaster}
  (1-\Omega_\chi)e^{4(\zeta_r-\zeta)}+\Omega_\chi
  e^{3(\zeta_\chi-\zeta)}=1~,
\end{equation}
where $\Omega_{\chi}=\bar\rho_{\chi}/(\bar\rho_r+\bar\rho_{\chi})$
is the dimensionless density parameter for the curvaton at the
decay moment. A necessary condition for the curvaton mechanism
working is based on the assumption that the fluctuation $\zeta_r$
seeded by the inflaton field is negligible. Then $\zeta$ and
$\zeta_\chi$ can be related order by order as
\begin{gather}\label{curvres123}
  \zeta_1=r\zeta_{\chi1}~,\qquad \frac{\zeta_2}{\zeta_1^2}=\frac{3}{2r}\left(1+\frac{gg''}{g'^2}\right)-r-2~,
  \nonumber\\
  \frac{\zeta_3}{\zeta_1^3}=\frac{9}{4r^2}\frac{g\left(3g'g''+gg'''\right)}{g'^3}
   -\frac{9}{r}\left(1+\frac{gg''}{g'^2}\right)-\frac{9gg''}{2g'^2}+\frac{1}{2}+10r+3r^2~,
\end{gather}
where $r$ is the transfer efficiency defined in the Introduction.
Finally, we have the nonlinearity parameter $f_{NL}$ in a usual
curvaton model as follows,
\begin{eqnarray}\label{fnlusual}
  f_{NL}=\frac{5}{4r}\bigg(1+\frac{gg''}{g'^2}\bigg)-\frac{5}{3}-\frac{5r}{6}~.
\end{eqnarray}
Additionally, the non-Gaussian estimator of the third order
fluctuation can be simplified in the absence of nonlinear
evolution of the curvaton field between the Hubble exit and the
start of curvaton oscillation which yields $g''=g'''=0$,
\begin{eqnarray}\label{gnlusual}
  g_{NL}=\frac{25}{54}\bigg(-\frac{9}{r}+\frac{1}{2}+10r+3r^2\bigg)~.
\end{eqnarray}
Notice that Eq. \eqref{gnlusual} corresponds to the simplest example
where a canonical curvaton has a quadratic potential. In this
case, one obtains $g_{NL}\propto r^{-1}$ which indicates that the
trispectrum is more difficult to be detected in the simplest
curvaton model. However, if $g$ is of a nontrivial form, the
trispectrum $g_{NL}$ could scale as $r^{-2}$ in the small $r$
limit.

We would like to emphasize two key features of the above
calculation. Firstly, the curvature perturbation $\zeta$,
$\zeta_\chi$ and $\zeta_r$ studied in the above are gauge
invariant. Therefore, although we study their dynamics in the
gauge of spatial flat slice, their values are irrelevant to the
gauge choice. Secondly, in the sudden decay approximation, $\zeta$
is conserved right after the universe evolves through the curvaton
decay hypersurface and the curvature perturbation should be
calculated exactly on this hypersurface. This is because, if one
calculates at any earlier time, $\zeta$ is not conserved; if one
calculates at any later time, one can no longer use the
information stored in $\zeta_\chi$. Having the above comments in
mind, it is clear that if the curvaton decay surface is different
from the uniform total energy density surface, the non-Gaussian
features are also changed. This is the right topic we shall
investigate in the next subsection.

\subsection{Curvaton decay at uniform curvaton density}\label{sec:2.2}

In this subsection, we phenomenologically assume that curvaton
decay happens on a uniform curvaton density slice. The realization
of this assumption on a brane is discussed in the next subsection.

If curvaton decay does not happen on the uniform total energy density
slice $\rho=\bar\rho(t_{\rm dec})$, instead, happens on the uniform
curvaton energy density slice
\begin{equation}\label{chislice}
  \rho_\chi=\bar\rho_\chi(t_{\rm dec})~,
\end{equation}
we can calculate the relation between $\zeta$, $\zeta_r$ and
$\zeta_\chi$ on the uniform curvaton density
slice\footnote{Another approach to calculating curvature
perturbation can be chosen on the slice of uniform total density
at some moment $t_T$ after curvaton decay. At that moment, we have
the total energy density
$\rho_{tot}(t_T)=\rho_{r,\chi}(t_T)e^{4({\zeta_{\chi}}_2-\delta{N}_T)}
+\rho_{r}(t_T)e^{4({\zeta_{r}}_T-\delta{N}_T)}$ and the curvature
perturbation $\zeta_T=\delta{N}_T$, where the subscript ``$T$"
denotes the time $t_T$. However, after curvaton decay the universe
has already entered a radiation dominant stage and thus is always
adiabatic which yields a conserved curvature perturbation
$\zeta_T=\zeta(t_{dec})$ at large scales. Therefore, the result
will not change if we compute the generation of primordial
perturbation at the slice of curvaton decay as shown in the main
text. We are grateful to Misao Sasaki for pointing out this
approach. }. In this case we apply Eq. \eqref{deltaN} on the slice
\eqref{chislice} and thus have
\begin{equation}
  \zeta_\chi=\delta N~.
\end{equation}
For radiation, we have
\begin{equation}\label{zetar}
  \zeta_r=\zeta_\chi+\frac{1}{4}\log\left(\frac{\rho_r}{\bar\rho_r}\right)~,
\end{equation}
and from Eq. \eqref{zetar}, we can solve $\rho_r$ as
\begin{equation}\label{rhor}
  \rho_r=\bar\rho_re^{4(\zeta_r-\zeta_\chi)}~.
\end{equation}
For the total energy density, we have
\begin{equation}\label{zetatot}
  \zeta=\zeta_\chi+\frac{1}{4}\log\left(\frac{\rho_r+3(\bar\rho_\chi+\bar P_\chi)/4}{\rho_r+3(\bar\rho_\chi+\bar P_\chi)/4}\right)~,
\end{equation}
in which we have applied the relation $d\tilde\rho=d\tilde\rho_r$,
since on the slice of uniform curvaton density $\rho_\chi$ is a
constant and can be absorbed into the background.

Inserting Eq. \eqref{rhor} into Eq. \eqref{zetatot}, we have
\begin{equation}
  \left(1-\frac{1-3w}{4}\Omega_\chi\right)e^{4(\zeta-\zeta_\chi)}=(1-\Omega_\chi)e^{4(\zeta_r-\zeta_\chi)}+\frac{3(1+w)}{4}\Omega_\chi~,
\end{equation}
where we have introduced an EoS parameter $w\equiv \bar P_\chi /
\bar\rho_\chi$ for the curvaton. This equation should be used to
replace Eq. \eqref{totalenergymaster} for doing the subsequent
calculation when the curvaton decay slice has uniform curvaton
energy density.

Again, we consider the limit of negligible $\zeta_r$, as is
assumed in a usual curvaton scenario. Of course, one could
generalize the calculation into the case of mixed perturbations of
inflaton and curvaton, however, this is beyond the scope of the
current paper. Thus the relation between $\zeta_\chi$ and
$\delta\chi_*$ is in principle not modified compared with the last
subsection. However, if we want to consider general EoS for
curvaton, the relation in the spatial flat slice (analog to Eq.
\eqref{deltaNchi}) becomes
\begin{equation}\label{deltaNchig}
  \rho_\chi=\bar\rho_\chi e^{3(1+w)\zeta_\chi}~.
\end{equation}
When $\delta\chi_*$ is Gaussian, to expand the above equations
order by order, we have the following generic relation for linear
curvature perturbation:
\begin{equation}\label{linearDBI}
  \zeta_1=\tilde{r}\zeta_{\chi1}~,
\end{equation}
where we have introduced a generalized transfer efficiency as
\begin{eqnarray}\label{tilder}
\tilde{r}\equiv\frac{3(1+w)\Omega_{\chi}}{4+(-1+3w)\Omega_{\chi}}~,
\end{eqnarray}
which encodes the information of the EoS of the curvaton field.

To proceed, we solve the curvature perturbation order by order
again and derive the non-Gaussian fluctuations as follows,
\begin{align}\label{fnlDBI}
  \frac{\zeta_2}{\zeta_1^2}=\frac{3(1+w)}{2\tilde{r}}\bigg(1+\frac{gg''}{g'^2}\bigg)+\frac{1-3w}{\tilde{r}}-4 ~,
\end{align}
\begin{align}\label{gnlDBI}
  \frac{\zeta_3}{\zeta_1^3} = \frac{9(1+w)^2}{4\tilde{r}^2}\frac{g(3g'g''+gg''')}{g'^3}
    +\frac{9(1+w)(1-3w)}{2\tilde{r}^2}\frac{gg''}{g'^2}
    +\frac{(5-3w)(1-3w)}{2\tilde{r}^2} \nonumber\\
    -\frac{18(1+w)}{\tilde{r}}\bigg(1+\frac{gg''}{g'^2}\bigg)
    -\frac{12(1-3w)}{\tilde{r}}+32 ~.
\end{align}
Substituting the above results into the non-Gaussian estimators
(\ref{estimator}), we are able to obtain the nonlinearity
parameters of local shape.

Very interestingly, large trispectrum can be generated even in the
case of a constant $g$. This conclusion is quite different from
that obtained in the previous subsection. Another thing worthy to
note is the generalized transfer efficiency parameter $\tilde{r}$
contains a factor $1+w$ in its expression. As a consequence, when
$w\rightarrow-1$, the value of $\tilde{r}$ can be strongly
suppressed and thus the amplitude of non-Gaussian fluctuations can
also be enhanced. This result indicates that the non-Gaussianity
could be amplified if there exists a secondary inflation. This
property was earlier discovered in several specific curvaton
models \cite{Huang:2008zj} and \cite{Cai:2009hw}, in the case that
curvaton decay still occurs on the slice of uniform total energy
density. For example, in \cite{Huang:2008zj} a secondary inflation
was achieved in the limit of $n\rightarrow0$ in their model; while
in \cite{Cai:2009hw} the similar background solution was obtained
due to the survival of a brane with a light mass term. In our
work, however, we point out clearly that the mechanism of
enhancing the non-Gaussian fluctuations due to a secondary
inflation is a generic feature in a curvaton model if the field
decays on the slice of uniform curvaton density as we analyzed.

\subsection{Application to models of non-relativistic curvaton branes}

The decay mechanism in the previous subsection has interesting
application in the case of brane inflation. This is because a
single D-brane is a BPS object, which is stabilized by
supersymmetry. The decay of a D-brane is typically expected to be
through the collision process with an anti D-brane.

We first consider the case when the curvaton is the position
modulus of a D3 brane\footnote{One should be careful in imposing
initial conditions to a curvaton brane. Usually one can impose
initial condition in two ways in curvaton scenario. One
possibility is that the curvaton originally sits at the minimum of
the potential. If the total inflationary e-folding number is much
larger than 60, the random walk of curvaton will finally set the
initial condition for curvaton for the last 60 inflationary
e-folds. Another possibility is to impose the initial condition
without the above random walk assumption. For brane inflation, one
can show that the quantum fluctuation during one Hubble time is
typically smaller than the classical motion \cite{Chen:2006hs}. If
this conclusion also applies for curvaton, one should expect only
the latter possibility for brane curvaton initial condition. On
the other hand, our scenario is motivated by string theory. As
inflation in string theory usually needs some fine-tuning
\cite{finet}, one expects the e-folding number is not much larger
than 60. This also supports the latter possibility. We thank
Xingang Chen for discussion on this point.}. The annihilation of a
curvaton D3 and an anti D3 takes place when the distance between
the curvaton D3 and the anti-D3 is of order string length. After
that, the potential of the brane-anti-brane system becomes
tachyonic, and the system decays into lower dimensional solitons.

In the framework of moduli stabilization, it is natural to assume
that the position modulus of the D3 is stabilized on the tip of the
deformed warped throat. This indicates that the open string modes
characterizing the position of the anti-D3 are very massive, and
so the anti-D3 is not allowed to move and even its fluctuations are
strongly depressed. In this case, the relative position between
the curvaton D3 and the anti-D3 is described by the position of
the curvaton D3, up to a constant. Thus the curvaton decay
surface becomes the uniform curvaton energy density slice, instead
of the uniform total energy density slice.

When the curvaton D3 moves non-relativistically, the brane
position can be described by a scalar field with a canonical
kinetic term. The most natural curvaton decay mechanism is that
the D3 moves towards the anti-D3 and then annihilates into a
system of tachyon condensate\cite{Cline:2002it} without an
oscillation period. In this case, the EoS of the position modulus
evolves from $w\simeq-1$ into $w\simeq0$ gradually as happened in
tachyon cosmology\cite{Gibbons:2002md}. Thus from Eqs.
\eqref{fnlDBI} and \eqref{gnlDBI}, the non-Gaussianities can be
enhanced during the occurrence of a secondary tachyonic inflation
with $w\simeq-1$ which will dilute other matter components in a
few e-folds; otherwise, we can also obtain large non-Gaussianities
due to a small value of $\Omega_\chi$ if the curvaton brane decays
into a pressureless tachyon condensate directly.

Now we consider another curvaton brane model achieved by a probe
anti-D3 with angular motion at the bottom of a warped throat with
approximate isometries\cite{Zhang:2009gw}. As usual, if a KS
throat is isometric, its warp factor is independent of the angular
coordinates. However, in a general case, there are some
corrections which make the warp factor dependent on angular
coordinates. For example, since the compact C-Y manifold cannot
have exact continuous isometries, the isometries of the bulk must
be broken when the finite throat is glued on this bulk, and then
the warp factor dependents on angles. Moreover, a nonperturbative
effect which stabilizes the K\"ahler moduli could bring a potential
for the branes\cite{Berg:2004ek, Baumann:2006th, DeWolfe:2007hd}.
The form of this potential depends on the precise embedding of the
wrapped branes. For a general embedding which does not admit
supersymmetric vacua on the tip, this potential could become
dependent on angles as well. Namely, an estimate of the
nonperturbative effect which is dominated by the warp factor in
the case of the Kuperstein embedding\cite{Kuperstein:2004hy} of
the D7-brane gives rise to a mass term for the angular degrees of
freedom, which is determined by the minimal radial location
reached by the D7 and the deformation parameter of the conifold.

In the model of curvaton brane involving angular degrees of
freedom, one may choose the flattest angular direction along the
rotation of the curvaton field. Since in this model the curvaton
brane is acted by an anti-D3, its radial coordinate is almost
fixed on the tip of the throat as explained above. In the limit of
a non-relativistic anti-D3, one could finely tune the model
parameters to allow the curvaton anti-D3 to rotate slowly so that
it can survive during inflation. After inflaton decay, the anti-D3 is
able to oscillate due to the existence of a nearly quadratic
potential in angular space.

In the above, we have introduced two explicit realizations of the
curvaton field in string compactification, i.e., by the distance
of a light D3 brane in the throat, as well by the flattest angular
degree of freedom of anti-D3 stabilized at the bottom of the
warped throat. For a light D3 brane, it decays only when it
arrives at the tip of the warped throat and annihilates with an
anti-D3; while for the angular degree of freedom of an anti-D3, it
decays only when the anti-D3 rotates to the position which is
parallel to others. In these cases, the value of the curvaton
field is almost the only parameter which determines the
annihilation of curvaton brane, and thus the curvaton decay. In
this sense, the inflaton energy density is not relevant to the
curvaton decay. Therefore, the curvaton decay slice is the uniform
curvaton field value slice, and thus the uniform curvaton energy
density slice.

Finally, we would like to point out that our calculation of
curvature perturbation on uniform curvaton density decay
hyper-surface is not limited to curvaton brane, but is generically
applicable to any other models in which the curvaton decays in the
uniform curvaton density slice. For example, our calculation could
be applicable for the curvaton scenarios in which the curvaton
decays from a narrow band preheating. In this preheating type of
curvaton scenario, curvaton decays each time when its field value
crosses zero, which corresponds to a uniform curvaton energy
density slice instead of a uniform total energy slice. Our method
also has significant implications if applied to the curvaton
scenario in the frame of bouncing cosmology\footnote{Some
observational signatures of bouncing cosmology were extensively
studied in the literature, such as curvature
perturbations\cite{Cai:2008qw, Cai:2008ed},
non-Gaussianities\cite{Cai:2009fn}, and constraints from current
observations\cite{Cai:2008qb}.}. However, since these scenarios
are beyond the scope of the current work, we would like to leave
the detailed analysis on these issues in future projects.

\section{The model of a relativistic curvaton brane}

In this section we extend the approach to studying the generation
of non-linearities on the uniform curvaton density slice into the
case involving a sound speed parameter. This can be obtained in an
example of a relativistic curvaton brane model. Specifically, in
the frame of warped compactifications, we consider a system
constructed by a number of D3-branes in a background metric field
with negligible covariant derivatives of the field strengths and
assume that these branes are falling into the warped throats, this
system can be described by a DBI action which has a stringy
origin.

\subsection{Background analysis}

We phenomenologically consider a double field inflation model with
its action in form of
\begin{equation}
  {\cal L}
  = \sqrt{-g} \left\{X-V(\phi)-\frac{1}{f}\sqrt{1-2fY}-W(\chi)\right\}~,
\end{equation}
where
\begin{equation}
  X\equiv -\frac{1}{2}\partial^\mu\phi\partial_\mu\phi~,\quad Y\equiv -\frac{1}{2}\partial^\mu\chi\partial_\mu\chi ~,\quad
  f\equiv\frac{1}{h^4 T_3}~.
\end{equation}
We shall find a solution in which $\phi$ behaves as inflaton,
which dominates the energy density during inflation, and $\chi$
behaves as curvaton, which can be realized by a mobile probe brane
in a warped throat during inflation.

To assume the curvaton $\chi$ is moving relativistically, we can
obtain a small sound speed defined by
\begin{equation}\label{relat}
  c_s\equiv\sqrt{1-2fY} \ll 1~,
\end{equation}
at the limit of $2fY\rightarrow1$. The continuous equation of
$\chi$ leads to
\begin{equation}
  \dot\chi \simeq -\frac{\sqrt{1-f\dot\chi^2}}{3H}W'~.
\end{equation}
One can solve $\dot\chi^2$ from this equation as
\begin{equation}
  2Y=\dot\chi^2=\frac{W'/(3H)}{1+fW'/(3H)}~.
\end{equation}
To make this equation consistent with the inequality
\eqref{relat}, the following condition is expected to be satisfied
in the model,
\begin{equation}\label{nece1}
  fW'/(3H) \gg 1~.
\end{equation}
Another necessary condition is that the assumption that inflaton
dominates the energy density:
\begin{equation}\label{nece2}
  V\gg W~.
\end{equation}
Making use of Eq. \eqref{nece2} and the Friedmann equation
$3M_p^2H^2\simeq V$, Eq. \eqref{nece1} can be rewritten as
\begin{equation}\label{nece3}
  f^2 W'^2M_p^2 \gg 3V \gg 3W~.
\end{equation}
Assuming that the curvaton potential is a mass term
$W=m^2\chi^2/2$, we need
\begin{equation}\label{nece4}
  f^2 m^2 M_p^2 \gg \frac{3}{2}~.
\end{equation}
If this condition is satisfied, we could have certain parameter
space to let $\chi$ behave as a relativistically moving curvaton.
Theoretically, this relation could be possible in condition of
that the warp factor $h$ is finely tuned to be sufficiently small
by adjusting the position of the curvaton brane in the throat. In
this case a sufficiently small warp factor yields a large value of
$f$ and therefore the field $\chi$ is able to slowly roll down
along the potential even this potential is not enough flat.
Consequently, $\chi$ is able to stay at sub-Planckian regime
safely, so that the backreaction on the warped throat could be
controllable.

Now let us examine the dynamics after inflation. Depending on the
choice of $f$ and $m$, as well as the brane position, the curvaton
might keep to stay in the relativistic regime, or slow down and
approach to a standard kinetic term. Practically, if the curvaton
keeps in the relativistic regime until it decays, there is no
oscillation stage of curvaton, and a secondary inflation driven by
curvaton is possible when the curvaton start to dominate. On the
other hand, when the curvaton becomes non-relativistic before its
decay, one can either have a stage of secondary inflation or
curvaton oscillation.

\subsection{Perturbation}

In a relativistic version of curvaton brane scenario, the
calculation of perturbation is the same as that of a single scalar
field with generalized kinetic terms in the slow roll
approximation. The only change is to write $\zeta_\chi$ as the
perturbation variable instead of $\zeta$. In this section, we
shall first calculate the three point function explicitly, which
shows the equivalence with single field DBI inflation. After that,
we directly apply the result of four point function of DBI
inflation to a relativistic curvaton brane scenario.

In the curvaton scenario, one can neglect the perturbation of the
inflaton field. As the curvaton field is subdominant during
inflation, one can ignore the perturbation of the metric. The
second order action of the curvaton takes the form
\begin{equation}
  S_2(\chi)\supseteq\int d^4 x \frac{a^3}{2}\left\{ \frac{\dot{\delta\chi}^2}{c_s^3}-\frac{(\partial_i\delta\chi)^2}{a^2c_s} \right\}~,
\end{equation}
in which the next-to-leading order terms have been neglected due
to a suppression of the slow roll condition.

We use $\zeta_\chi$ to denote the curvature perturbation in the
uniform curvaton energy density slice. To leading order in $c_s$,
one can write
\begin{equation}
  \zeta_\chi=-\frac{H\delta\chi}{\dot\chi}~.
\end{equation}
We assume that the curvaton field is not directly coupled to the
inflaton field (except for the coupling via gravity). In this
case, as shown in \cite{Lyth:2004gb}, $\zeta_\chi$ is a conserved
quantity on super-Hubble scales. At the moment of Hubble-crossing,
one can solve out $\delta\chi_*=\frac{H_*}{2\pi}$, and so the
corresponding power spectrum is given by
\begin{equation}
  P_{\zeta}=\tilde{r}^2P_{\zeta_\chi}=\tilde{r}^2\frac{H^{4}}{4\pi^2\dot\chi^2}.
\end{equation}

\subsubsection{Bispectrum}

To expand the action into cubic order, we obtain the dominant part
of the action as follows,
\begin{equation}\label{eq:s3curvaton}
  S_3(\chi)\supseteq\int d^4 x \frac{a^3}{2}\left(\frac{1}{c_s^2}-1\right)\left\{ \frac{\dot{\delta\chi}^3}{c_s^3\dot\chi}-\frac{\dot{\delta\chi}(\partial_i\delta\chi)^2}{a^2c_s\dot\chi}
   \right\}~.
\end{equation}
The initial correlation function of $\chi$ can be calculated using
the time dependent perturbation theory.
Firstly, the mode function of the perturbation can be written as
\begin{equation}
  u_k=\frac{H}{\sqrt{2k^3}}\left(1+ikc_s\tau\right)e^{-ikc_s\tau}~,\quad u'_k=\frac{H\sqrt{k}}{\sqrt{2}}c_s^2\tau e^{-ikc_s\tau}~.
\end{equation}
One can show that the first term in
the third order action leads to a contribution to the three-point
function as
\begin{equation}
  \langle \zeta_\chi^3 \rangle \supset \frac{3H^8}{2\dot\chi^4}\left(\frac{1}{c_s^2}-1\right) \frac{(2\pi)^3\delta^3(\sum {\bf k_i})}{K_3^3k_1k_2k_3}~,
\end{equation}
and the second term in the third order action leads to a
contribution to the three-point function as
\begin{equation}
  \langle \zeta_\chi^3 \rangle \supset - \frac{H^8}{4\dot\chi^4}\left(\frac{1}{c_s^2}-1\right)\frac{(2\pi)^3\delta^3(\sum {\bf k_i})({\bf k}_2\cdot{\bf k}_3)}{k_1k_2^3k_3^3}
  \left(\frac{1}{K_3}+\frac{k_2+k_3}{K_3^2}+\frac{2k_2k_3}{K_3^3}\right)+2{~\rm perm.}~.
\end{equation}
Summing them up, we have
\begin{equation}\label{bichi}
  \langle \zeta_\chi^3 \rangle = \left(\frac{1}{c_s^2}-1\right)P_{\zeta_\chi}^2{(2\pi)^7\delta^3(\sum {\bf k_i})}
  \left(-\frac{1}{K_3}\sum_{i>j}k_i^2k_j^2+\frac{1}{2K_3^2}\sum_{i\neq j}k_i^2k_j^3+\frac{1}{8}\sum_i k_i^3\right)~.
\end{equation}
Substituting Eq. \eqref{linearDBI} into Eq. \eqref{bichi}, the
three-point function for curvature perturbation $\zeta$ takes the
form
\begin{align}\label{bizeta}
  \langle \zeta^3 \rangle = \frac{1}{\tilde r}\left(\frac{1}{c_s^2}-1\right)P_{\zeta_\chi}^2{(2\pi)^7\delta^3(\sum {\bf k_i})}
  \left(-\frac{1}{K_3}\sum_{i>j}k_i^2k_j^2+\frac{1}{2K_3^2}\sum_{i\neq j}k_i^2k_j^3+\frac{1}{8}\sum_i k_i^3\right)
  +\langle\zeta^3\rangle_{\rm loc}~,
\end{align}
where $K_3=k_1+k_2+k_3$, and $\langle\zeta^3\rangle_{\rm loc}$ is
given by Eqs. \eqref{shapeB} and \eqref{estimator}. The leading
order non-Gaussianity is a combination of equilateral shape and
local shape. In the $c_s\rightarrow 0$ limit, the shape becomes
completely equilateral. When $c_s$ is not very close to zero, the
contribution $\langle\zeta^3\rangle_{\rm loc}$ is also important.

\subsubsection{Trispectrum}

For the trispectrum, as shown in \cite{Chen:2009bc,
Arroja:2009pd}, the correlation function includes a scalar
propagation part and a contact interaction part. The detailed
shape is complicated. Instead of providing the complete form of the
lengthy shape function ({as shown in the appendix}), we would like
to study the representative shape here. The trispectrum can be
approximated by
\begin{align} \label{trizeta}
  \langle\zeta^4\rangle = \frac{1}{\tilde{r}^2}(2\pi)^9P_\zeta^3\delta^3(\sum\bk_i)\frac{\alpha}{c_s^4}\frac{k_1^2k_2^2k_3^2k_4^2}{(K_4)^5}
   +\langle\zeta^4\rangle_{\rm loc}+\langle\zeta^4\rangle_{\rm cross}
  ~,
\end{align}
where $K_4=k_1+k_2+k_3+k_4$, and $\alpha$ is a constant in order
of unity. The local part $\langle\zeta^4\rangle_{\rm loc}$ is
given by Eqs. \eqref{shapeT} and \eqref{estimator}. The cross
correlation can be calculated as
\begin{align}
  \langle\zeta^4\rangle_{\rm cross}=&\frac{3}{5}f_{NL}^{\rm loc}\langle (\zeta*\zeta)(\bk_1) \z{2}\z{3}\z{4} \rangle + 3{~\rm perm.}
  \nonumber\\&
  =\frac{3}{5}f_{NL}^{\rm loc}P(k_2)\langle \zeta(\bk_1-\bk_2)\z{3}\z{4} \rangle + 11{~\rm perm.}~,
\end{align}
and $\langle \zeta(\bk_1-\bk_2)\z{3}\z{4} \rangle$ is calculated
using Eq. \eqref{bichi}.

\section{Non-Gaussian features of curvaton fluctuations in specific examples}

In the above section we have studied the generation of curvature
perturbation up to third order. In order to obtain a much
intuitive insight of these non-Gaussian features, one still needs
to evaluate the nonlinearity parameters $f_{NL}$ and $g_{NL}$ in
explicit cases. As we have introduced in Section 2.3, there exist
two plausible evolution trajectories for the curvaton brane after
inflaton decay. We study these possibilities respectively in the
following. In these specific examples, we generically consider the
curvaton potential is almost quadratic around the moment of
curvaton decay so that we are able to apply the formalism
developed in Section 2. Therefore, when we compute the
non-Gaussian fluctuations at curvaton decay, the probe brane has
already become non-relativistic as a canonical field. However, one
may notice that the sound speed parameter $c_s$ still exists in
the following computation, since it was inherited from the
Hubble-crossing of curvaton fluctuation during inflation. On the
other hand, when the probe brane is still moving relativistically
near curvaton decay, one can use the techniques developed in
\cite{Cai:2009hw} to generalize the calculation of this Section.

\subsection{Case of $w \rightarrow -1$}

We first consider the case that near the curvaton decay, the
curvaton brane has EoS $w \rightarrow -1$. The theoretical
realization of this scenario was discussed in Ref.
\cite{Cai:2009hw}. If we consider the curvaton brane is falling
into the AdS-like throat and neglect the backreaction of the
branes upon the background geometry, the warp factor usually takes
the form of $f(\chi)=\lambda/\chi^4$. Note that, the amplitude of
the curvature perturbation depends on the term $\dot\chi$ of which
the lower bound corresponds to its value near the bottom of the
warp throat. The warp factor at the infrared (IR) end of the
throat is given by $h_{IR}\simeq\exp\left(-\frac{2\pi
K}{3g_sM}\right)$ where $g_s$ is the string coupling, $M$ and $K$
are the RR charge on the $S_3$ cycle and NS-NS charge on the dual
cycle respectively. Therefore, at the relativistic limit, one
obtains
\begin{eqnarray}
\dot\chi_{IR}=\chi_{IR}^2\sqrt{\frac{1-c_s^2}{\lambda}}\simeq\sqrt{T_3}h_{IR}^2~,
\end{eqnarray}
at the IR cutoff.

The key parameter related to the generation of primordial
curvature perturbation in the curvaton model is the perturbation
transfer efficiency $\tilde{r}$ defined in Eq. (\ref{tilder}).
From the definition, one learns that this parameter is determined
by the EoS $w$ and density parameter $\Omega_{\chi}$ of the
curvaton. If the curvaton brane decays when its energy density is
still sub-dominate, then there could be a double suppression on
the value of $\tilde{r}$ because of a combined effect of
$w\rightarrow-1$ and $\Omega_{\chi}\rightarrow0$.

However, if the curvaton does not decay until it become dominate,
the universe enters a secondary inflation driven by the curvaton.
The energy density of the curvaton will catch up with that of
radiation decayed from inflaton in the first several e-folds. As a
consequence, it yields $\Omega_\chi\rightarrow 1$  at the moment
of curvaton decay. Note that Eq. \eqref{tilder} can be rewritten
as
\begin{equation}
  \tilde r=\frac{3\Omega_\chi}{4(1-\Omega_\chi)/(1+w)+\Omega_\chi}\simeq \frac{3}{4(1-\Omega_\chi)/(1+w)+1}~,
\end{equation}
and thus one concludes that the order of magnitude of $\tilde r$
is determined by the ratio between $1-\Omega_\chi$ and $1+w$.
Namely, when $(1-\Omega_\chi)/(1+w)\gg 1$, one has $\tilde r \ll
1$; and when $(1-\Omega_\chi)/(1+w)\ll 1$, one has $\tilde r
\simeq 3$.

The correlation functions are given by Eqs. \eqref{bizeta} and
\eqref{trizeta} respectively. One can decompose the non-Gaussian
estimator into local and equilateral ones by comparing their
squeezed limit. The local components for $f_{NL}$ and $g_{NL}$ are
given by Eqs. \eqref{fnlDBI}, \eqref{gnlDBI} and
\eqref{estimator}. The equilateral components for $f_{NL}$ and
$g_{NL}$ are given by \cite{Chen:2006nt, Chen:2009bc}
\begin{equation}
  f_{NL}^{\rm eq}=\frac{35}{108\tilde
    r}\left(\frac{1}{c_s^2}-1\right)~,\quad
  g_{NL}^{\rm eq}\simeq \frac{0.542}{c_s^4\tilde r^2}~,
\end{equation}
which implies the equilateral non-Gaussian fluctuations can be
doubly enhanced if both $\tilde{r}$ and $c_s^2$ have small values.

\subsection{Case of $w \rightarrow 0$}

After inflation, the curvaton brane will arrive at the tip of the
throat. At that moment, either the open string modes of the
curvaton brane lead to a matter-like tachyon condensate with
$w\simeq-c_s^2 \rightarrow 0$ as studied in Ref.
\cite{Li:2008fma}, or a non-perturbatively moduli stabilization
brings a quadratic potential for the angular modes of the brane
which gives rise to a non-relativistic oscillation of the curvaton
with $w\simeq0$ in average and
$c_s\simeq1$\cite{Kobayashi:2009cm}. Consequently, we have two
plausible ending processes of the curvaton brane which yields
different results for non-Gaussianities.

\subsubsection{Local Non-Gaussianity}

According to the analysis performed in Section \ref{sec:2.2}, we
learn that the non-Gaussian fluctuations in local limit are
insensitive to the sound speed $c_s$ but mainly depend on the
parameter of transfer efficiency $\tilde{r}$. When $w\simeq0$ the
form of $\tilde{r}$ coincides with the usual transfer efficiency
$r$. Therefore, following Eqs. (\ref{fnlDBI}) and (\ref{gnlDBI}),
the non-Gaussian estimators (\ref{estimator}) give the
nonlinearity parameter of local shape as follows,
\begin{eqnarray}
 f_{NL} \simeq \frac{5}{4r}\left(1+\frac{gg''}{g'^2}\right)+\frac{5}{6r}-\frac{10}{3}~,
\end{eqnarray}
when the curvaton decay takes place on the slice of uniform
curvaton density.

In the limit of $r\rightarrow1$ when the curvaton dominates the
universe before it decays, the nonlinearity parameter is given by,
\begin{eqnarray}
 f_{NL}\rightarrow-\frac{5}{4}\left(1-\frac{gg''}{g'^2}\right)~,
\end{eqnarray}
of which the value is negative and sizable if $g''$ is negligible.
On the other hand, we have a local non-Gaussianity in the limit of
$r\ll1$ which takes the form,
\begin{eqnarray}
 f_{NL} \rightarrow
 \frac{5}{4r}\left(\frac{5}{3}+\frac{gg''}{g'^2}\right)~,
\end{eqnarray}
which is amplified by a small $r$. We notice that the above result
is consistent with the result obtained in a usual canonical
curvaton model only in the case of $r\rightarrow1$, but deviates
from the usual model in the case of $r\rightarrow0$ with a factor
of order $O(1)$.

In the absence of higher nonlinear evolution of the $\chi$ field
between the Hubble exit and the decay moment, we would have
$g'''=0$ for simplicity. In this case, the third order
perturbation gives
\begin{eqnarray}
 g_{NL} \simeq
 \frac{25(5-8r)}{24r^2}\frac{gg''}{g'^2}+\frac{125}{108r^2}-\frac{125}{9r}+\frac{400}{27}~.
\end{eqnarray}
Again we consider two limited cases and list the results as
follows,
\begin{eqnarray}
 g_{NL} &\rightarrow& \frac{25}{12}(1-\frac{3gg''}{2g'^2})\bigg|_{r\rightarrow1}~,\\
        &\rightarrow&
        \frac{125}{108r^2}\left(1+\frac{9gg''}{2g'^2}\right)\bigg|_{r\rightarrow0}~.
\end{eqnarray}
The result in the first limit is in agreement with that of a usual
curvaton model. However, the latter one is quite different since
its form is enhanced by order of $r^{-2}$ even when $g''=0$, but
in a canonical curvaton model with a quadratic potential we have
$g_{NL}\propto r^{-1}$ when $g''$ is negligible.

\begin{figure}[htbp]
\includegraphics[scale=0.75]{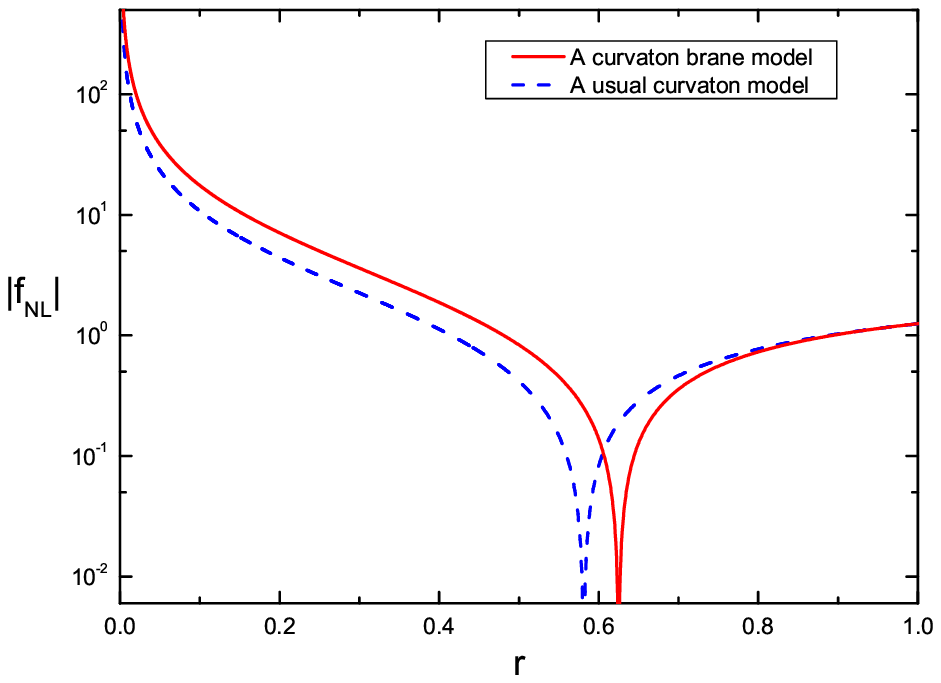}
\includegraphics[scale=0.75]{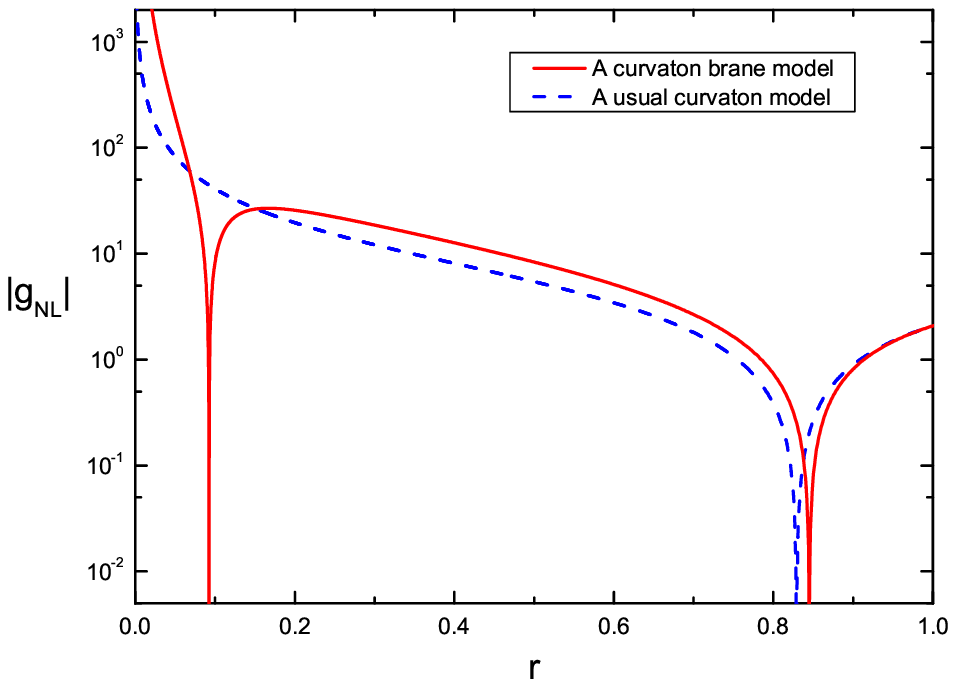}
\caption{Absolute values of the nonlinearity parameters $f_{NL}$
and $g_{NL}$ of local shape as a function of the transfer
efficiency $r$ in the model of curvaton brane and the usual
curvaton scenario. Here we assume that $g(\chi)$ is linear for
simplicity.} \label{fig:ng}
\end{figure}

In order to understand the above analytic study more clearly, we
compare the nonlinearity parameters $f_{NL}$ and $g_{NL}$ of local
shape in the model of curvaton brane with the results obtained in
a usual scenario in Figure. \ref{fig:ng}. In this figure, we
numerically plot the nonlinearity parameters as functions of the
transfer efficiency $r$, and consider the simplest curvaton model
without nonlinear evolution of the $\chi$ field between the Hubble
exit and the moment of curvaton decay which gives a linear
function of $g(\chi)$. The red solid curve represents for the
non-Gaussianity of local shape when the curvaton field decays
on the hypersurface of uniform curvaton density; while, the blue
dash line denotes the non-Gaussianity of local shape when the
curvaton decay takes place on the slice of uniform total energy
density. One notices that these curves coincide at the
limit of $r\rightarrow1$, but possess different behavior when $r$
is less than unity. The result for $g_{NL}$ is more manifest
than for $f_{NL}$.

\subsubsection{Equilateral non-Gaussianity}

Now we study the non-Gaussian fluctuation of equilateral shape in
the case of $w\rightarrow0$. Since we have pointed out that the
sound speed $c_s^2$ can be either very small in the case of
tachyon condensate or approach unity when the curvaton obtains
a quadratic potential, we phenomenologically treat the sound speed
as a free parameter in the following. Again, we obtain the
equilateral components for $f_{NL}$ and $g_{NL}$ in form of
\begin{equation}
  f_{NL}^{\rm eq}=\frac{35}{108 r}\left(\frac{1}{c_s^2}-1\right)~,\quad
  g_{NL}^{\rm eq}\simeq \frac{0.542}{c_s^4 r^2}~.
\end{equation}
This result also indicates that both small $r$ and $c_s^2$ are
able to amplify the equilateral non-Gaussianity in the model of
curvaton brane with $w\simeq0$.

From above analysis,  if the curvaton brane moves
non-relativistically during inflation with $c_s^2\simeq1$, there
are $f_{NL}\propto1/r$ and $g_{NL}\propto1/r^2$ for both the local
and equilateral shapes, but the amplitudes of local shape is much
larger than equilateral ones; however, if the curvaton brane moves
relativistically with $c_s^2\simeq0$ in inflationary stage, the
amplitudes of $f_{NL}$ and $g_{NL}$ could be greatly amplified by
the order of $1/c_s^2$ and $1/c_s^4$ respectively. As a
consequence, one concludes that when the mobile curvaton brane is
non-relativistic, the shape of non-Gaussianity is mainly of local
type with its corresponding size being different from that in the
canonical curvaton model, but becomes of equilateral type in the
case of a relativistic curvaton brane.

\section{Conclusion}

In this paper, we have calculated the nonlinear primordial
curvature perturbation in the curvaton scenario using a
generalized $\delta{N}$ formalism.
If curvaton decay does not occur on the slice of uniform total
energy density but on the slice of uniform curvaton density
instead, we find that the dynamics of the nonlinear fluctuations
behaves different from the usual scenario. Specifically, we
consider a model of curvaton brane which can provide a theoretical
realization of the curvaton decay we expected. In the frame of
this model, we have presented a full analysis on the sizes and
shapes of its bispectra and trispectra respectively. We arrive at
an important conclusion that the generation of nonlinearities is
sensitive to the mechanism of curvaton decay, which is mainly
determined by the EoS $w$, the density occupation $\Omega_{\chi}$
of the field $\chi$ at curvaton decay and its sound speed $c_s$
during inflation. Explicitly speaking, our results show that the
nonlinearity parameter characterizing the second order
perturbation $f_{NL}$ is proportional to $1/\tilde{r}$ in which
the transfer efficiency relies on the EoS $w$ and the density
occupation $\Omega_{\chi}$ of the curvaton. For the equilateral
shape, the size of $f_{NL}$ is amplified by a large value of the
factor $1/c_s^2$, and therefore when the curvaton brane moves
relativistically the shape of non-Gaussianity is mainly of
equilateral type; however, at the squeeze limit the nonlinearity
parameter $f_{NL}$ decouples from the sound speed and its size is
qualitatively consistent with the result obtained in a usual
canonical curvaton model. Moreover, for the third order
fluctuations, we find the nonlinearity parameter $g_{NL}$ is
enhanced by the factor $1/c_s^4$ in equilateral shape and
decouples in local limit, and in the simplest example its size is
proportional to $1/\tilde{r}^2$ which is quite different from the
usual scenario.

As an end, we would like to highlight the importance of our study
in this paper. The mechanism of curvaton decay after inflation
could determine the relation of the primordial curvature
perturbation and the decay hypersurface. This process is rather
robust and should be considered in any specific curvaton model,
and the example of curvaton brane analyzed in the current work is
a good illustration to emphasize its importance. Additionally, the
signatures of curvaton decay imprinted on the primordial curvature
perturbation could provide a new window to explore the combination
of early universe physics and astronomical observations.

\medskip
\section*{Acknowledgments}
We wish to thank Neil Barnaby, Xingang Chen, Damien Easson, Andrew
Frey, Takeshi Kobayashi, Shinji Mukhoyama, Misao Sasaki, and
Xinmin Zhang for valuable comments on our work. Y.F.C. thanks the
Institute for the Physics and Mathematics of the Universe and the
Research Center for the Early Universe at the University of Tokyo,
Tokyo University of Science, and the Yukawa Institute for
Theoretical Physics at Kyoto University for their hospitality when
this work was finalized. The work of Y.F.C. is supported in part
by the Arizona State University Cosmology Initiative. Y.W. thanks
the Canadian Institute for Theoretical Astrophysics for
hospitality. The work of Y.W. is supported in part by NSERC and an
IPP postdoctoral fellowship.

\section*{Appendix: Full shape function of the trispectrum}

The shape of trispectrum for a relativistic brane curvaton is the
same as that when the brane behaves as an inflaton. For
completeness, here we present the full shape function of the
trispectrum, which is calculated in \cite{Chen:2009bc}. As usual,
we define \bea c_s^2 &\equiv& \frac{P_{,X}}{P_{,X}+2X P_{,XX}} ~,
\cr \Sigma &\equiv& X P_{,X} + 2X^2 P_{,XX} ~, \cr \lambda
&\equiv& X^2 P_{,XX} + \frac{2}{3} X^3 P_{,XXX} ~, \cr \mu
&\equiv& \half X^2 P_{,XX} + 2X^3 P_{,XXX} + \frac{2}{3} X^4
P_{,XXXX} ~. \eea

\begin{figure}
\centering
\includegraphics{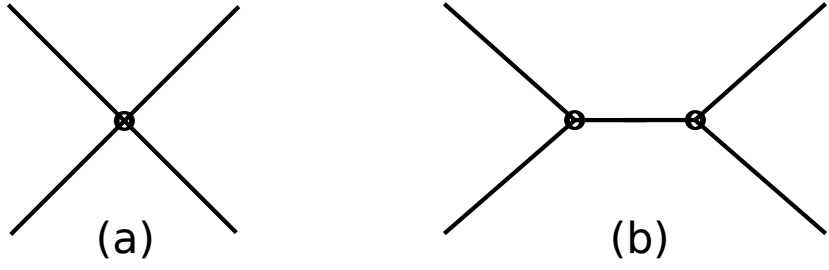}
\caption{\label{fig:4pt} Two classes of contributions to the
trispectrum. The contribution (a) is denoted by ``contact
interaction'' diagram, and the contribution (b) is denoted by
``scalar exchange'' diagram. }
\end{figure}

There are two classes of contributions to the trispectrum, namely,
the contact interaction and scalar exchange diagrams. They are
shown in Fig. \ref{fig:4pt}(a) and \ref{fig:4pt}(b), respectively.
Using the method of time dependent perturbation theory, the former 
can be calculated from the forth order
action, and the latter can be calculated by applying the third
order action twice.

We will skip the lengthy calculation, and refer the readers to
\cite{Chen:2009bc} for details. It turns out that the trispectrum
for $\zeta_\chi$ can be written as \bea \langle \zeta_\chi^4
\rangle = (2\pi)^9 P_{\zeta_\chi}^3 \delta^3(\sum_{i=1}^4 \bk_i)
\prod_{i=1}^4 \frac{1}{k_i^3} ~ \CT(k_1,k_2,k_3,k_4,k_{12},k_{14})
~, \eea where $\CT$ has six components: \bea \CT &=& \left(
\frac{\lambda}{\Sigma} \right)^2 T_{s1} + \frac{\lambda}{\Sigma}
\left( \frac{1}{c_s^2}-1 \right) T_{s2} + \left( \frac{1}{c_s^2}-1
\right)^2 T_{s3} + \left( \frac{\mu}{\Sigma} -
\frac{9\lambda^2}{\Sigma^2} \right) T_{c1} \cr &&+
\left(\frac{3\lambda}{\Sigma} - \frac{1}{c_s^2} +1 \right) T_{c2}
+ \left( \frac{1}{c_s^2} -1 \right) T_{c3} ~, \label{shapesum}
\eea where $T_{c1}$, $T_{c2}$ and $T_{c3}$ are contributions from
contact interaction, which are given by \bea T_{c1}\equiv 36
\left( \frac{\mu}{\Sigma} - \frac{9 \lambda^2}{\Sigma^2} \right)
\frac{\prod_{i=1}^4 k_i^2}{K^5} ~, \label{CT_c1} \eea

\bea
T_{c2}\equiv -\frac{1}{8} \left(\frac{3\lambda}{\Sigma} -
\frac{1}{c_s^2} +1 \right)
\frac{k_1^2k_2^2 (\bk_3 \cdot \bk_4)}{K^3}
\left[ 1+ \frac{3(k_3+k_4)}{K} + \frac{12k_3k_4}{K^2} \right]
+ {\rm 23~perm.} ~;
\eea

\bea T_{c3}&\equiv&\frac{1}{32} \left( \frac{1}{c_s^2} -1 \right) \frac{(\bk_1
  \cdot \bk_2)(\bk_3 \cdot \bk_4)}{K} \left[ 1+ \frac{\sum_{i<j} k_i k_j}{K^2} +
  \frac{3k_1k_2k_3k_4}{K^3} (\sum_{i=1}^4 \frac{1}{k_i} ) + 12
  \frac{k_1k_2k_3k_4}{K^4} \right] \cr &&+ ~{\rm 23 ~ perm.} ~. \label{CT_c3}
\eea $T_{s1}$, $T_{s2}$ and $T_{s3}$ come from interaction with
scalar exchange. As shown in Eq. (\ref{eq:s3curvaton}), there are
two dominant terms in the curvaton third order action. One can
show that the coupling coefficients of these two terms are of
order $\lambda/\Sigma$ and $\frac{1}{c_s^2}-1$, respectively.
$T_{s1}$ denotes the contribution that both interaction vertices
in Fig. \ref{fig:4pt}(b) have coupling $\lambda/\Sigma$, and thus
is proportional to $(\lambda/\Sigma)^2$. $T_{s2}$ denotes the
contribution which is proportional to
$\lambda/\Sigma(\frac{1}{c_s^2}-1)$. Additionally, $T_{s3}$
denotes the contribution with two powers of $\frac{1}{c_s^2}-1$.

The summation $T_{s1}+T_{s2}+T_{s3}$ are given by a summation of
the following contributions from Eq. \eqref{aa1_1} to Eq.
\eqref{bb23_4}:
\bea \frac{9}{8} \left( \frac{\lambda}{\Sigma}
\right)^2 k_1^2 k_2^2 k_3^2 k_4^2 k_{12}
\frac{1}{(k_1+k_2+k_{12})^3 M^3} + {\rm 23~ perm.} ~.
\label{aa1_1} \eea

\bea
\frac{9}{4} \left( \frac{\lambda}{\Sigma} \right)^2
k_1^2 k_2^2 k_3^2 k_4^2 k_{12} \frac{1}{M^3}
\left( \frac{6M^2}{K^5} + \frac{3M}{K^4} + \frac{1}{K^3} \right)
+ {\rm 23~ perm.} ~.
\label{aa23_1}
\eea

\bea -\frac{3}{32}
\frac{\lambda}{\Sigma} \left( \frac{1}{c_s^2} - 1 \right) (\bk_3
\cdot \bk_4) k_{12} k_1^2 k_2^2 \frac{1}{(k_1+k_2+k_{12})^3}
F(k_3,k_4,M) + 23~ {\rm perm.} ~. \label{ab1_1} \eea

 \bea -\frac{3}{16}
\frac{\lambda}{\Sigma} \left( \frac{1}{c_s^2} -1 \right) (\bk_{12}
\cdot \bk_4) \frac{k_1^2 k_2^2 k_3^2}{k_{12}}
\frac{1}{(k_1+k_2+k_{12})^3} F(k_{12},k_4,M) + 23~ {\rm perm.} ~.
\label{ab1_2} \eea

\bea - \frac{3}{16}
\frac{\lambda}{\Sigma} \left( \frac{1}{c_s^2} -1 \right) (\bk_3
\cdot \bk_4) k_1^2 k_2^2 k_{12} ~ G_{ab}(k_3,k_4) +23~{\rm perm.} ~.
\label{ab23_1} \eea

\bea - \frac{3}{8}
\frac{\lambda}{\Sigma} \left( \frac{1}{c_s^2} -1 \right) (\bk_{12}
\cdot \bk_4) \frac{k_1^2 k_2^2 k_3^2}{k_{12}} ~ G_{ab}(k_{12},k_4)
+23~{\rm perm.} ~. \label{ab23_2} \eea
 \bea - \frac{3}{16}
\frac{\lambda}{\Sigma} \left( \frac{1}{c_s^2} -1 \right) (\bk_1
\cdot \bk_2) k_3^2 k_4^2 k_{12} ~ G_{ba}(k_1,k_2) +23~{\rm perm.} ~.
\label{ba23_1} \eea

 \bea \frac{3}{8}
\frac{\lambda}{\Sigma} \left( \frac{1}{c_s^2} -1 \right) (\bk_2
\cdot \bk_{12}) \frac{k_1^2 k_3^2 k_4^2}{k_{12}} ~
G_{ba}(-k_{12},k_2) +23~{\rm perm.} ~. \label{ba23_2} \eea
 \bea \frac{1}{2^7} \left(
\frac{1}{c_s^2} -1 \right)^2 (\bk_1 \cdot \bk_2)(\bk_3 \cdot \bk_4)
k_{12} ~ F(k_1,k_2,k_1+k_2+k_{12}) F(k_3,k_4,M) + {\rm 23~perm.} ~.
\label{bb1_1} \eea
 \bea \frac{1}{2^5}
\left( \frac{1}{c_s^2} -1 \right)^2 (\bk_1 \cdot \bk_2)(\bk_{12}
\cdot \bk_4) \frac{k_3^2}{k_{12}} ~ F(k_1,k_2,k_1+k_2+k_{12})
F(k_{12},k_4,M) + {\rm 23~perm.} ~. \eea
 \bea -\frac{1}{2^5} \left(
\frac{1}{c_s^2} -1 \right)^2 (\bk_{12} \cdot \bk_2)(\bk_{12} \cdot
\bk_4) \frac{k_1^2 k_3^2}{k_{12}^3} ~ F(k_{12},k_2,k_1+k_2+k_{12})
F(k_{12},k_4,M) + {\rm 23~perm.} ~. \eea
 \bea \frac{1}{2^6} \left(
\frac{1}{c_s^2} -1 \right)^2 (\bk_1 \cdot \bk_2)(\bk_3 \cdot \bk_4)
k_{12} ~ G_{bb}(k_1,k_2,k_3,k_4) + {\rm 23~perm.} ~. \eea
 \bea \frac{1}{2^5} \left(
\frac{1}{c_s^2} -1 \right)^2 (\bk_1 \cdot \bk_2)(\bk_{12} \cdot
\bk_4) \frac{k_3^2}{k_{12}} ~ G_{bb}(k_1,k_2,k_{12},k_4) + {\rm
23~perm.} ~. \eea
 \bea -\frac{1}{2^5} \left(
\frac{1}{c_s^2} -1 \right)^2 (\bk_{12} \cdot \bk_2)(\bk_3 \cdot
\bk_4) \frac{k_1^2}{k_{12}} ~ G_{bb}(-k_{12},k_2,k_3,k_4) + {\rm
23~perm.} ~. \eea
 \bea -\frac{1}{2^4} \left(
\frac{1}{c_s^2} -1 \right)^2 (\bk_{12} \cdot \bk_2)(\bk_{12} \cdot
\bk_4) \frac{k_1^2k_3^2}{k_{12}^3} ~ G_{bb}(-k_{12},k_2,k_{12},k_4)
+ {\rm 23~perm.} ~. \label{bb23_4} \eea
where
\bea
\bk_{12}=\bk_1+\bk_2 ~, ~~~
M=k_3+k_4+k_{12}~, ~~~ K=k_1+k_2+k_3+k_4~,
\eea
and $F$, $G_{ab}$, $G_{ba}$, $G_{bb}$ are defined as follows:
\bea
&&F(\alpha_1,\alpha_2,m)
\cr
&\equiv& \frac{1}{m^3}
\left[ 2\alpha_1\alpha_2 + (\alpha_1+\alpha_2)m +m^2 \right] ~,
\label{Fdef}
\\
\nonumber \\
&&G_{ab}(\alpha_1,\alpha_2)
\cr
&\equiv&
\frac{1}{M^3K^3}
\left[ 2\alpha_1 \alpha_2 +(\alpha_1+\alpha_2)M +M^2 \right]
\cr
&+&
\frac{3}{M^2K^4}
\left[ 2\alpha_1 \alpha_2 +(\alpha_1+\alpha_2)M \right]
+ \frac{12}{MK^5} \alpha_1\alpha_2 ~,
\\
\nonumber \\
&&G_{ba}(\alpha_1,\alpha_2)
\cr
&\equiv&
\frac{1}{M^3K} + \frac{1}{M^3K^2}(\alpha_1+\alpha_2+M)
+\frac{1}{M^3K^3}
\left[ 2\alpha_1\alpha_2 + 2(\alpha_1+\alpha_2)M +M^2 \right]
\cr
&+&
\frac{3}{M^2K^4}
\left[ 2\alpha_1\alpha_2 + (\alpha_1+\alpha_2)M \right]
+ \frac{12}{MK^5}\alpha_1\alpha_2 ~,
\label{Gba}
\\
\nonumber \\
&&G_{bb}(\alpha_1,\alpha_2,\alpha_3,\alpha_4)
\cr
&\equiv&
\frac{1}{M^3 K} \left[ 2 \alpha_3 \alpha_4 + (\alpha_3+\alpha_4)M +
M^2 \right]
\cr
&+& \frac{1}{M^3 K^2}
\left[ 2 \alpha_3\alpha_4(\alpha_1 + \alpha_2)
+ \left( 2 \alpha_3 \alpha_4 +
(\alpha_1 + \alpha_2)(\alpha_3+\alpha_4) \right) M
+ \sum_{i=1}^4 \alpha_i M^2 \right]
\cr
&+& \frac{2}{M^3 K^3}
\left[ 2 \prod_{i=1}^4 \alpha_i +
\left(2\alpha_3\alpha_4(\alpha_1+\alpha_2) +
\alpha_1\alpha_2(\alpha_3+\alpha_4) \right) M
+ \sum_{i<j} \alpha_i \alpha_j M^2 \right]
\cr
&+& \frac{6}{M^2 K^4} \left(\prod_{i=1}^4 \alpha_i \right)
\left( 2+M \sum_{i=1}^4 \frac{1}{\alpha_i} \right)
+ \frac{24}{MK^5} \prod_{i=1}^4 \alpha_i
~.
\label{Gbb}
\eea
Note that in $G_{ab}$, $G_{ba}$ and $G_{bb}$, the $K$ and $M$ are
defined as $K=k_1+k_2+k_3+k_4$ and $M=k_3+k_4+k_{12}$ (which changes
correspondingly in permutations), but not in terms
of $\alpha_i$'s.

\end{document}